\documentclass[preprintnumbers,amsmath,11pt,amssymb,floatfix,superscriptaddress,nofootinbib]{article}

\def\mytitle#1{\setcounter{equation}{0}
	\setcounter{footnote}{0}
	\begin{flushleft}\Large\textbf{#1}\end{flushleft}
	\vspace{0.25cm}}
\def\myname#1{\leftline{{\large #1}}\vspace{-0.13cm}}
\def\myplace#1#2{\small\begin{flushleft}\textit{#1}\\
		\texttt{#2}\end{flushleft}}

\usepackage{graphicx}
\usepackage{dcolumn}
\usepackage{color}
\usepackage{bm}
\usepackage{amsmath}
\usepackage{amsfonts}
\usepackage{amssymb}

\input epsf

\begin{document}

\mytitle{Modified Hawking temperature and entropic force: a prescription in FRW model}
\myname{Saugata Mitra\footnote{saugatamitra20@gmail.com}}
\myplace{Department of Mathematics, Jadavpur University, Kolkata 700032, West Bengal, India.}
\author{Subhajit Saha\footnote {subhajit1729@gmail.com}}
\myplace{Department of Mathematics, Jadavpur University, Kolkata 700032, West Bengal, India.}
\author{Subenoy Chakraborty\footnote {schakraborty.math@gmail.com}}
\myplace{Department of Mathematics, Jadavpur University, Kolkata 700032, West Bengal, India.}{}

\begin{abstract}
The idea of Verlinde that gravity is an entropic force caused by information changes associated with the positions of material bodies, is used in the present work for the FRW model of the Universe. Using modified Hawking temperature, the Friedmann equations are derived on any horizon. For the validity of the first law of thermodynamics (i.e., Clausius relation) it is found that there is modification of Bekenstein entropy on the horizon. However, using equipartition law of energy, Bekenstein entropy is recovered. 
\end{abstract}


In 1967, Sakharov \cite{sa1,sa2,sa3,sa4} first raised the question whether gravity should be regarded as a fundamental interaction in nature. According to him, space-time background emerges as a mean field approximation of some underlying microscopic degrees of freedom and as a result there is induced gravity.

Few years back, Verlinde \cite{v} again raised this issue considering holographic principle and the equipartition rule. He was able to derive newtonian law of gravitation as an entropic force and he claimed that gravity is not a fundamental interaction, rather an emergent phenomenon based on the statistical behavior of microscopic degrees of freedom encoded on a holographic screen. According to him, the force of gravity is not ingrained in matter itself, rather an extra physical effect emerging from the interplay of mass, time, space and information. In particular, gravity is treated as an entropic force caused by changes in the information associated with the positions of material bodies. Also using this idea of entropic force with the Unruh temperature \cite{u}, he derived Newton's law of force (2nd law). In this context, it is worthy to mention that combining the equipartition law of energy for the horizon degrees of freedom with the thermodynamical relation $S=\frac{E}{2T}$, (S= entropy, T= temperature of the horizon and E is the active gravitational mass $\cite{p}$), Padmanabhan $\cite{p3}$ was also able to derive newtonian gravitational law. One should also note that there were earlier attempts to construct microscopic models of space-time  \cite{p2,d,m}.

It should be mentioned that Jacobson $\cite{ja}$ and Padmanabhan $\cite{p3}$ introduced the idea that gravity is not a fundamental interaction, rather an emergent phenomenon based on the statistical behavior of an ensemble of microscopic degrees of freedom belonging to the general idea of holography (general holographic principle). On the other hand, the approach of Verlinde $\cite{v}$ is a much more restrictive one, making considerably stronger assertions- the differences being both technical and conceptual. For example, in Jacobson and Padmanabhan scenarios, one intends to derive the Einstein field equations, not individual forces on individual particles which in contrast to the Verlinde scenario. Further, in Verlinde's proposals, (with an Unruh-like temperature and entropy related to "holographic screens") one is unavoidably, forced into a more general 'thermodynamic force' scenario with multiple intensive and extensive thermodynamic variables with 'entropies' that are negative and the 'temperatures' which are positive for attractive forces and negative for repulsive forces.

From the cosmological view point, using the ideas of information and holography, entropy and temperature, a different approach $\cite{r1}$ is adopted to explain present observations without the introduction of dark energy (DE). It is assumed that the Universe is bounded by a horizon (apparent / event) and there should be temperature and entropy associated with the horizon similar to Hawking temperature and Bekenstein entropy of a black hole ( BH) horizon.

Recently, it has been found that the temperature on the event horizon is not $\frac{1}{2\pi R_E}$, in analogy to that of the apparent horizon (i.e., $\frac{1}{2\pi R_A}$) \cite{cai,nm,c,ca1}, rather the temperature on any horizon ($R_h$) is given by \cite{c,ca1}
\begin{equation}
T_h^{(m)}=\left(\frac{R_h}{2\pi R_A^2}\right)\frac{\hbar c}{\kappa_B}
\end{equation}
and is termed as modified Hawking temperature on the horizon, where $\kappa_B$ is the Boltzmann's constant, $\hbar$ is the reduced Planck constant and c is the velocity of light. Note that it coincides with the usual Hawking temperature on the apparent horizon.

Further, the temperature of the horizon leads to the usual entropic force and the resultant acceleration ($a_{horizon}$) of the horizon is given by the Unruh relation $\cite{u}$,
$$a_{horizon}=\frac{2\pi \kappa_B }{\hbar}T^{(m)}_A=cH \sim 10^{-9}m/s^2,$$
where 
 $$T^{(m)}_A=\frac{c \hbar}{\kappa_B}\left(\frac{H}{2\pi}\right) \sim 3 \times 10^{-30} K$$
is the modified Hawking temperature on the apparent horizon, with $H$ as the Hubble parameter.
 
One may note that the above cosmic acceleration is consistent with observation.

 Further, due to entropic force at the horizon, there is an outward pull which creates the appearance of a DE component. Also the pressure exerted by the entropic force is negative and is close to the recently measured DE negative pressure.

It should be noted that similar to late-time accelerated expansion, inflation can also be explained using the concept of entropic force $\cite{r3}$. In fact, holographic principle and entropy which combine to dark entropic geometry is considered as the source of accelerating phases of the Universe. For various difficulties and loopholes with entropic forces in gravity and cosmology, one may refer to \cite{r4,r5,r6,r7} and references therein.

In this letter, we shall derive the following results of universal thermodynamics for the FRW model of the Universe based on the idea of entropic force:\\
i) Friedmann equations from entropic force using modified Hawking temperature on any horizon.\\
ii) Modified entropy on the horizon for the validity of the first law of thermodynamics on the horizon.\\
iii) Bekenstein entropy on the horizon from equipartition law of energy.\\\\\\
{\bf Derivation of Friedmann equations}\\

According to Verlinde \cite{v}, suppose a compact spatial region V has a compact boundary $\partial V$ which can be considered as the holographic screen. If the boundary $\partial V$ is taken as the surface of a sphere with physical radius R (= area radius) then the number of bits on the screen is written as $\cite{sh}$
\begin{equation}
N=\frac{A }{c_1 L_P^2},
\end{equation}
where A is the area of the screen , $c_1$ is a numerical factor whose value will be determined later and $L_P=\sqrt{\frac{G \hbar}{c^3}}$ is the Planck length.

In thermodynamics, fluctuations of the area elements in the hot Rindler horizon play a crucial role $\cite{p3}$. In analogy with this, we shall assume that each cell of area $c_1 L_P^2$ contributes an energy $\frac{\kappa_B}{2}T$ and is in accordance with the standard thermodynamic equipartition law.
Then the total energy on the screen is
\begin{equation}
E= \frac{1}{2} N \kappa_B T,
\end{equation}
where T is the temperature on the screen chosen as the modified Hawking temperature (i.e., $T=T_h^{(m)}$).

Verlinde \cite{v} then considered the relativistic energy-mass relation, i.e.,
\begin{equation}
E=Mc^2
\end{equation}
to relate the above total energy on the screen to the mass M that would emerge in the compact spatial region V enclosed by the boundary screen $\partial V$. Now, mass in the spatial region V can be expressed as
\begin{equation}
M=\int_V dV (T_{\mu \nu} u^\mu u^\nu )=\frac{4}{3}\pi R^3 \rho,
\end{equation}
where $\rho=~ T_{\mu \nu} u^\mu u^\nu $ is the energy density measured by a co-moving observer and $R=ar$ is the area radius in the FRW model (a is the scale factor).

If we now consider the compact boundary $\partial V$ which acts as the holographic screen to be a boundary of the FRW model of the space-time, then its temperature is the modified Hawking temperature given by equation (1) ($R_h$ is the area radius of the horizon). Now eliminating N, E, M and T (i.e., $T_h^{(m)}$) from equations (1)-(5), we have
\begin{equation}
\frac{1}{R_A^2}=\frac{4\pi G c_1 \rho}{3},
\end{equation}
on the holographic screen at the horizon. As for FRW model $R_A=\frac{1}{\sqrt{H^2+\frac{k}{a^2}}}$, so from the above relation, we have the Friedmann equation:
\begin{equation}
H^2+\frac{k}{a^2}=\frac{8\pi G \rho}{3},
\end{equation}
where $a$ is the scale factor of the Universe and $k$ is the curvature scalar. Here we have chosen $c_1=2$.

Further, if perfect fluid is chosen as the matter source in the Universe, then the stress-energy tensor has the form
\begin{equation}
T_{\mu \nu}=(\rho+p)u_\mu u_\nu +pg_{\mu \nu},
\end{equation}
where $\rho,~\textrm{and}~ p$ are the energy density and thermodynamic pressure of the fluid respectively,  and $u_\mu$ is a unit time-like vector. Due to the presence of the pressure, the total mass $M=\rho V$ bounded by the holographic screen $\partial V$ can no longer be conserved, rather a variation in the total mass is bounded by the work done, by the pressure i.e., $dM=-pdV$ which leads us to the continuity equation
\begin{equation}
\dot{\rho}+3H(\rho+p)=0.
\end{equation}
Now differentiating the first Friedmann equation (7) with respect to time and eliminating $\dot{\rho}$ from (9), we have the second Friedmann equation, namely
\begin{equation}
\dot{H}-\frac{k}{a^2}=-4\pi G (\rho+p).
\end{equation}
Thus using the idea of entropic force and modified Hawking temperature, it is possible to derive the Friedmann equations on any horizon.\\\\\\
{\bf First law of thermodynamics and modified entropy on any horizon:}\\\\
If we choose $c=\hbar=G=\kappa_B=1$, then from the equipartition law (3) (using (2)), taking differential of both sides we obtain
\begin{equation}
dE_h=\frac{1}{c_1}d\left(\frac{R_h^3}{R_A^2}\right).
\end{equation}
So for the validity of the first law of thermodynamics (i.e., Clausius relation), we must have
\begin{equation}
T_h^{(m)}dS_h=dE_h=\frac{1}{c_1}d\left(\frac{R_h^3}{R_A^2}\right),
\end{equation}
where $S_h$ is the entropy on the surface of the horizon (i.e., holographic screen). Using (1), we obtain
\begin{equation}
S_h^{(m)}=\frac{2\pi}{c_1}\int R_h^2 d\left\lbrace\left(\frac{R_h^3}{R_A^2}\right)\right\rbrace,
\end{equation}
where $S_h^{(m)}$ is termed as the modified entropy on the horizon. Thus we see that using the idea of entropic force and modified Hawking temperature, we always have first law of thermodynamics but at the cost of non-Bekenstein entropy (given by equation (13)) on the horizon. It should be noted that for apparent horizon the above modified entropy becomes
\begin{equation}
S_A^{(m)}=\frac{1}{c_1}S_{BA},
\end{equation}
where $S_{BA}$ is the usual Bekenstein entropy on the apparent horizon. Further, if the horizon is assumed to be close to the apparent horizon so that
\begin{equation}
R_h=R_A+\epsilon,
\end{equation}
then with the first order in $\epsilon$, we have
\begin{equation}
S_H^{(m)}=S_A^{(m)}-\frac{2\pi \epsilon}{c_1}R_A
\end{equation}
Thus the first order correction to the modified entropy is linear to the radius of the horizon.\\\\
{\bf Bekenstein entropy from equipartition law of energy}\\\\
On the other hand, if instead of considering the first law of thermodynamics, we use the equipartition law of energy, then what will be the form of the entropy function?

 According to Padmanabhan $\cite{p3}$, if a patch of horizon is divided into N (given in equation (2)) microscopic cells and each shell has $c_2$ internal states then the total number of microscopic states is $c_2^N$ and hence the entropy of the screen (i.e., horizon) is given by \cite{cai,nm,c,ca1}\\
\begin{equation}
S=N\ln c_2 =\frac{4 \ln c_2}{c_1}\left(\frac{A c^3}{4 L_P^2}\right)=\frac{4 \ln c_2}{c_1}S_B.
\end{equation}
Hence entropy on the holographic screen is proportional to the corresponding Bekenstein entropy.\\\\
{\bf Modified Hawking temperature versus Unruh temperature}\\\\
If the holographic screen (S) is defined as an equipotential surface then the four acceleration of a particle close to the holographic screen is given by $\cite{c2}$
\begin{equation}
a^\mu=-\nabla^\mu \phi,
\end{equation}
where $\phi$ is the newtonian potential. According to Padmanabhan $\cite{p}$, such acceleration is produced by the active gravitational mass \cite{p,c1} (known as Tolman-Komar mass) instead of total mass in the spatial volume. Then the Unruh temperature of the holographic screen corresponding to this acceleration is given by
\begin{equation}
T_U=-\frac{\hbar}{2\pi c \kappa_B}e^\phi n^\mu a_\mu = \frac{\hbar}{2\pi c \kappa_B}e^\phi \sqrt{(\nabla^\mu \phi)(\nabla_\mu \phi)}
\end{equation}
where $n^\mu=\frac{\nabla^\mu \phi}{\sqrt{(\nabla^\lambda \phi)(\nabla_\lambda \phi)}}$ is the unit normal vector on the holographic screen.
Now, in order to produce the above acceleration, Verlinde \cite{v} has chosen the energy of the screen 'S' as the Komar mass energy \cite{b,t,t1}, i.e.,
\begin{equation}
E_K=\frac{1}{2}\int_S T_UdN,
\end{equation}
where N is the number of bits stored on the holographic screen S.
The above equation can be interpreted as the energy equipartition rule on the screen. Assuming the temperature on the screen to be constant, we have
\begin{equation}
dE_K=\frac{1}{2}T_UdN=\frac{1}{2}T_U\frac{c^3}{c_1 G \hbar}dA,
\end{equation}
where we have used equation (2). As before, choosing $c_1=2$, we have the Clausius relation
\begin{equation}
dE_K=T_UdS_B
\end{equation}
on the horizon (holographic screen).

Thus for the Unruh temperature, choosing Komar mass density as the energy equipartition rule on the screen, the first law of thermodynamics holds on the holographic screen, with Bekenstein entropy on the screen.

Lastly, we shall have a comparative study of the above two temperatures (i.e., modified Hawking temperature and Unruh temperature) at any horizon (i.e., holographic screen) where for simplicity, we consider a radial co-moving observer on the screen. The Unruh temperature is given by
\begin{equation}
T_U=\frac{\hbar}{2\pi \kappa_B c}a_r
\end{equation}
with
\begin{equation}
a_r=-\frac{d^2R}{dt^2}=-\frac{\ddot{a}}{a}R,
\end{equation}
the acceleration of the observer. Hence for spatially flat FRW model,
\begin{equation}
\frac{T_U}{T_h^{(m)}}=q
\end{equation}
where $q=-\frac{\ddot{a}/a}{H^2}$ is the usual deceleration parameter.\\\\

In this letter, we have derived Friedmann equations on any horizon for the FRW Universe using Verlinde's concept of gravity as an entropic force. In reference  \cite{c3}, the authors have derived the Friedmann equations of a FRW Universe starting from the holographic principle together with the equipartition law of energy using Verlinde's concept of gravity as an entropic force and the Unruh temperature. On the other hand, in the present work, the Friedmann equations are derived choosing the holographic screen as the surface of any horizon of the FRW Universe, the modified Hawking temperature (on the horizon) as the temperature and the energy corresponding to the total mass bounded by the holographic screen (Cai et al. \cite{c3} had considered energy corresponding to the active gravitational mass (i.e., Tolman Komar mass)).
Then, using Verlinde's idea that a test particle approaching the holographic screen will cause an entropy variation and using equipartition law, the first law of thermodynamics is found to be valid but the entropy on the horizon is no longer Bekenstein in nature (Bekenstein entropy with a constant multiplicative factor at the apparent horizon) and modified entropy can be expressed in an integral form (see equation (13)). Here also we have used modified Hawking temperature on the screen.
However, instead of using the first law of thermodynamics on the holographic screen, if equipartition law of energy is considered, then entropy on the screen can be shown to be Bekenstein entropy provided the arbitrary parameter $c_1$ is related to the number of microscopic configuration of each cell by the relation $c_1=4\ln c_2$. As we have chosen $c_1=2$, so the number of internal states in each microscopic cell is chosen as $\sqrt{e}$.
Subsequently, if we restrict to local Rindler causal horizon and a uniformly accelerated frame \cite{p3}, it is possible to show the validity of the first law of thermodynamics assuming uniform temperature on the holographic screen (i.e., the surface of the horizon) and using equipartition of energy on the screen. Throughout the work we have chosen the arbitrary numerical factor $c_1$ to be 2 (compared to $c_1=1$ by most of the works in related areas) and we are able to derive the Friedmann equations from the idea of entropic force. Bekenstein entropy is derived from the equipartition law of energy by restricting the internal states of a microscopic cell. Also we have derived the first law of thermodynamics using Unruh temperature and choosing Komar mass density as the energy equipartition rule on the screen. The only bad feature is that starting from the first law of thermodynamics we are not able to derive Bekenstein entropy (with $c_1=2$) even at the apparent horizon rather we have a multiplicative constant. Further it should be noted that the numerical factor $c_1$ is related to the number of internal states in each microscopic cell by the relation $4\ln c_2=c_1$. Thus choosing $c_1=2$, the number of internal states is increased from $e^{\frac{1}{4}}$ (for $c_1=1$) to $e^{\frac{1}{2}}$. Also from (2), we see that the number of microscopic cells is reduced by half when $c_1$ changes from unity to $c_1=2$. Hence the total number of microscopic states remain same ($=e^{\frac{N}{4}}$) in each case. Thus it is not mandatory to have $c_1=1$, rather we must have $4\ln c_2=c_1$ and we have shown that the number of microscopic states remain same. This point has also been mentioned by Padmanabhan \cite{p3}.

Finally, it should be mentioned that although Verlinde derived the Einstein equations in a general static background with a time like Killing vector field, but they can as well be derived for a homogeneous FRW space-time. However, the choice of the parameter $c_1$ should be taken care of. For further work, it would be interesting to find the conditions for applicability of Unruh temperature and modified Hawking temperature and their relations with acceleration.
\section*{Acknowledgement}
The author S.M. is thankful to UGC for NET-JRF.
The author S.S. is thankful to UGC-BSR Programme of Jadavpur University for awarding Research Fellowship.
S.C. is thankful to UGC-DRS programme, Department of Mathematics, Jadavpur University.

\end{document}